# Diagnosing COVID-19 Pneumonia from X-Ray and CT Images using Deep Learning and Transfer Learning Algorithms

Halgurd S. Maghdid, Aras T. Asaad, Kayhan Zrar Ghafoor, Ali Safaa Sadiq, Muhammad Khurram Khan

*Abstract*—COVID-19 (also known as 2019 Novel Coronavirus) first emerged in Wuhan, China and spread across the globe with unprecedented effect and has now become the greatest crisis of the modern era. The COVID-19 has proved much more pervasive demands for diagnosis that has driven researchers to develop more intelligent, highly responsive and efficient detection methods. In this work, we focus on proposing AI tools that can be used by radiologists or healthcare professionals to diagnose COVID-19 cases in a quick and accurate manner. However, the lack of a publicly available dataset of X-ray and CT images makes the design of such AI tools a challenging task. To this end, this study aims to build a comprehensive dataset of X-rays and CT scan images from multiple sources as well as provides a simple but an effective COVID-19 detection technique using deep learning and transfer learning algorithms. In this vein, a simple convolution neural network (CNN) and modified pre-trained AlexNet model are applied on the prepared X-rays and CT scan images dataset. The result of the experiments shows that the utilized models can provide accuracy up to 98% via pre-trained network and 94.1% accuracy by using the modified CNN.

## I. INTRODUCTION

The novel COVID-19 was first reported in Wuhan city, Hubei Province of China, China in November 2019. A month later, the World Health Organization (WHO) announced that the virus can cause a respiratory disease with clinical presentation of cough, fever and lung inflammation. Although the COVID-19 emerged in China, it has now been identified in many other countries around the globe [1], [2]. On January 30, 2020, WHO announced this pandemic as public health emergency. This is not only due to its fast person-to-person spread but also because most infected people are not immune to it.

Halgurd S. Maghdid is with the Department of Software Engineering, Faculty of Engineering, Koya University, Kurdistan Region-F.R.Iraq. First.Last@koyauniversity.org.
Aras Asaad is with the Oxford Drug Design, Oxford Centre for Innovation, Oxford, OX1 1BY, UK.aras.asaad@oxforddrugdesign.com
Kayhan Zrar Ghafoor is with the Department of Software Engineering, Salahaddin University-Erbil, Iraq; School of Mathematics and Computer Science, University of Wolverhampton, Wulfruna Street, Wolverhampton, WV1 1LY, UK. kayhan@ieee.org.
Ali Safaa Sadiq is with the School of Mathematics and Computer Science, University of Wolverhampton, Wulfruna Street, Wolverhampton, WV1 1LY, UK. Ali.Sadiq@wlv.ac.uk.
Muhammad Khurram Khan is with the Centre of Excellence in Information Assurance, King Saud University, Riyadh, Saudi Arabia. mkhurram@KSU.EDU.SA

The COVID-19 spreads among people and many different species of animals such as camels, cattle, cats and bats. Early cases of COVID-19 at the epicenter in Wuhan had links with seafood and live animal markets, indicating animal-to-person transmission. The number of infected people has been growing significantly as a result of person-to-person transmission of this virus [3]. Thereafter, the WHO on March 11, 2020, when the number of confirmed cases reached 118,000 with more than 4000 deaths, announced that the novel COVID-19 outbreak as a pandemic [4]. Recently, Italy overtakes the China in number of fatalities.

COVID-19 and human coronaviruses are categorised under the family of Coronaviridae. These viruses infect people with moderate cold Middle East Respiratory Syndrome (MERS) or Severe Acute Respiratory Syndrome (SARS) [5]. SARS is also a viral respiratory disease caused by SARS-associated coronavirus (SARS-CoV), which was first reported in 2003 in Southern China and spread in many countries worldwide. Moreover, MERS virus cases were first announced in Saudi Arabia causing 858 deaths. Based on the analysis of virus genomes, this virus is believed to have originated in bats [6]. The clinical presentation of COVID-19 is complicated and could be manifested as fever, cough and severe headache. There are several techniques for COVID-19 detection including the Nucleic Acid Test (NAT) and Computed Tomography (CT) scan. The NAT is utilized to detect specific nucleic acid sequence and species of organism, predominately a virus or bacteria that causes disease in blood, tissue or urine. Although NAT and detection kits play significant roles in detecting COVID-19, CT scans remain the most effective and functional for detecting the severity and degree of the lung inflammation likely to be associated with COVID-19 [7]. The National Health Commission of China confirmed the inclusion of radiographic presentation of pneumonia for clinical diagnostic standard in Hubei Province [8], which assures the significance of CT scan images for the diagnosis of COVID-19 pneumonia severity.

Recently, a surge of COVID-19 patients has introduced long queues at hospitals for CT scan image examination. This leads to a serious risk of cross-infection with other patients and consequently overloads the medical system. Moreover, the number of radiologists is relatively far smaller than the number of patients which can lead to late detection and quarantine of infected people and less efficient treatment of patients [8]. As a result, for instance, recently in Italy, hospitals have had to give priority to people with a significant fever and shortness of breath over others with less severe symptoms [9].

The rapid spread of COVID-19 and the overwhelming demand for diagnosis has driven researchers to develop more intelligent, highly responsive and efficient diagnosis methods. One diagnosis method, handled by the radiologists, is the manual lung infection quantification. Further, there is an AI-based automated pneumonia diagnosis used to identify the density and volume of lesions and opacities of confirmed COVID-19 cases. Such algorithms are able to analyze the results of CT scan images in a short time in comparison to other existing methods [10].

Radiologists use chest CT scan images to follow up the confirmed cases ranging from early to critical stages [10]. The quick progression of lung infection requires multiple CT scan images Interpreting and analyzing these images can be a time consuming and daunting task, especially when manually quantifying the infected regions on the CT scan and X-ray images. Thus, there is a pressing need to develop an intelligent algorithm to accurately and automatically detect COVID-19 cases. Furthermore, it is required to build a complete and ready-to-experiment dataset for research community.

This research first reviews the state-of-the-art solutions to combat COVID-19. Then, we build a pre-processed and comprehensive dataset on X-rays and CT scan images from multiple sources, and provide an accurate COVID-19 detection algorithm using deep learning and transfer learning tools. Further, a modified CNN and AlexNet model as a pre-trained network are applied on the prepared X-rays and CT scan images datasets. After extensive experiments on both datasets, it is shown that the proposed model predicts COVID-19 diagnosis with high accuracy and low response time. To the best of our knowledge, there is no work in the open literature that uses both pretrained and CNN on both CT scan and X-ray images in identifying COVID-19.

From a technical standpoint, we summarise the most important contributions of this paper as follows:

1) We build a ready-to-use dataset of CT scan and X-ray images. For this purpose, we use multiple sources of those images in order to make the proposed model more realistic in detecting COVID-19 cases.
2) We propose a modified CNN model to accurately diagnose COVID-19 and hence delay the fast spread of the virus.
3) In order to accelerate and elevate the accuracy of diagnosis, we also propose a modified pre-trained deep learning model to detect COVID-19.

The rest of this paper is organized as follows. Section II provide the literature review on recent advances of developed AI systems for COVID-19 detection. This is followed by presenting an overview of the proposed approach and details of the designed algorithm. Finally, Section IV concludes the paper.

## II. BACKGROUND

As was stated earlier that Computed tomography (CT) scan method is the ideal way for diagnosing novel coronavirus (COVID19) pneumonia. The authors in [1] have conducted a research intended to build a diagnose system based on deep learning for identifying COVID-19 pneumonia. Their system works with high-resolution CT scan images in diagnosing COVID19, which supports radiologists with their work and helps to control the epidemic. It was interesting that they could manage to collect 46,096 anonymous images from 106 admitted patients, involving 51 cases of laboratory confirmed COVID-19 pneumonia. Besides, they have collected 55 control patients of other diseases admitted in Renmin Hospital of Wuhan University (Wuhan, Hubei province, China). All these CT scan images were prospectively collected to develop and train their proposed model as well as to evaluate and contrast the efficiency of radiologists versus COVID19 pneumonia with the performance of their model.

It is important to mention that their proposed deep learning model has exhibited an equivalent performance compared with expert radiologist. Moreover, it could also significantly improve the efficiency of radiologists while doing their clinical practice; which is crucially needed with such outbreaks situations when cases exponentially increasing. Hence, we can highlight here an argument, having such diagnosing system grips numerous potential to alleviate the burden of frontier radiologists in addition to advance the timely COVID19 diagnosis. Furthermore, obtaining an early detection of such highly contagious pandemic will further help in isolation and cure plans, and eventually assists countries in controlling and ending such epidemic [1].

There was also a very important claim highlighted in [1] on the importance of CT scan images in diagnoses the cases of COVID19, which can be considered a much quicker method in comparison with the conventional way using the nucleic acid detection. In additional to its efficiency in diagnosing the infection, it can estimate the level of severity of pneumonia [11]. It was reported that CT results were positive with all 140 laboratory-confirmed COVID-19 patients. Some of these positive cases, CT scan was able to recognize them even within their early stage, which reflects its effectiveness [12], [13]. On the other hand, the National Health and Health Commission of China has reported in their fifth version of COVID-19 diagnostic manual, the radiographic characteristics of pneumonia is integrated the clinical diagnostic standard in Hubei Province.

From the aforementioned discussion, the researchers in [1] were mainly focusing to achieve a model that could closely diagnosis COVID19 to the way radiologists do, but with shorter time. They could achieve a comparable performance to that of expert radiologist with 65% lesser time taken in diagnosing cases compared to in-clinic radiologist's time. Through, there are still points of improvements available for further enhancing their proposed model and the entire system to be personally accessible by users. This will enable patients as well as suspected cases to have their self-check system so they save more time and avoid direct contact that may lead for disseminating the virus to the specialists or nurses.

On the other hand, the authors in [10] have stated that till the time of writing their report, there is no such automatic toolkit to clinically quantify the level of COVID-19 infection for patients. For this motive, they have introduced a deep learning (DL)-based system to automatically segmenting and quantifying the infection spots of the COVID19 patients in addition to obtaining the entire view of lung to be extracted from chest CT scans. To perform image segmentation, a DL-based network named VB-Net was proposed. Authors in [10] have developed a modified 3-D convolutional neural network, which associations V-Net [14] along with the bottleneck structure presented in [15]. The adopted VB-Net is presented in Fig. 1, which entails of two routes, the dashed boxes highlighting the utilized bottleneck elements within their V-shaped network.

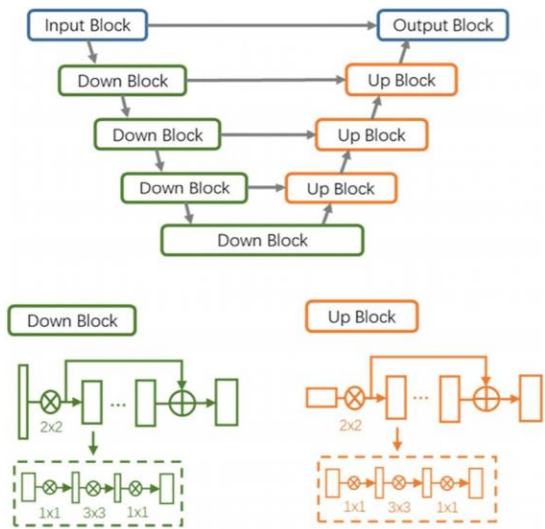

Fig. 1. Network structure in [10] for COVID-19 infection segmentation.

In contrast, Fig. 2 illustrating the proposed human-in-the-loop workflow strategy by the authors in [10]. Their DL-based segmentation has involved the "VB-Net" neural network to extract all COVID-19 infection areas out of CT scan images. They have trained their system with 249 COVID- 19 cases, and validated it with other 300 new COVID-19 cases. We can see that they have also introduced a human-in-the-loop (HITL) strategy to be used during training phase of their model to speedup process of manual definition of CT scan images. This strategy has helped radiologists during the process of tagging some footnotes on each case's scanned image. The proposed model was implemented and tested across the conventional way of diagnosing the COVID19 to find that the proposed system could reduce 4 minutes after 3 iterations of model updating, in contrast with the fully manual delineation; which regularly may take time varied between 1 to 5 hours.

In another attempt, Adrian Rosebrock in [16] has proposed a DL model for diagnoses COVID19 using Keras library and TensorFlow training platform. Using in total 50 images, which has been equally divided into 25 positive COVID-19 and other 25 negative X-rays images; they have build, train and validate their model. Out of the conducted experiments, their proposed model could diagnoses COVID-19 with average accuracy of 90-92%, which was applied on their testing set with 100% sensitivity and 80% specificity due to the limited data. As been stated in the published work in [10], the proposed model still has high potential for improvement, as it was trained and tested only with relatively small dataset. Besides, the model has to be trained and learn numerous patterns, which are not highly related to COVID-19 so that can broad its diagnoses knowledge system. It is also crucial highlighting that having a relatively robust CONVID19 diagnoses system could be achieved via a multi-modal, which processes multi factors such as patient vitals, population density, geographical location and some others. Therefore, having a diagnoses system relying only on X-ray images will not be that reliable to ending such high risk epidemic of COVID-19 [16].

## III. THE PROPOSED APPROACH

This section discusses the design of the proposed CNN approaches used to identify COVID-19 cases and details of proposed CNN model implementations. The proposal includes two main algorithms which a simple CNN architecture and a transfer learning algorithm, namely AlexNet. CNN algorithms are working by extracting relevant features through a series of convolutional layers followed by fully connected neural layers. There are many types of deep learning algorithms ranging from convolution neural networks (CNN) to the recurrent neural networks (RNN). The CNN could be applied on these solutions when the data are retrieved in a spatial domain such as image processing applications [12]. But the RNN is working on the concept of re-using the output of each layer as an input for the next layers. Further, the RNN is compatible with those applications which are getting sequential data such text [17] and signal reading measurements [18]. While, the transfer learning is the concept of re-using a pre-trained network and transfer the learned model into a new model. The new models can also take new additional training data and modified neural layers [19]. In this study, for the purpose of diagnosing COVID-19 cases, a CNN architecture (as a deep learning algorithm) and modified AlexNet network (as a transfer learning algorithm) have been utilized.

**Proposed CNN architecture:** To detect COVID-19 cases, we designed a very simple CNN model that consists of only one convolutional layer that constitute of 16 filters followed by batch normalization, rectified linear unit (ReLU), two fully connected layers, SoftMax and a classification layer. We adopted 'glorot' to initialize the weights and cross-entropy used as a loss function in the classification layer. Full details of our proposed CNN model is depicted in Fig. 3.

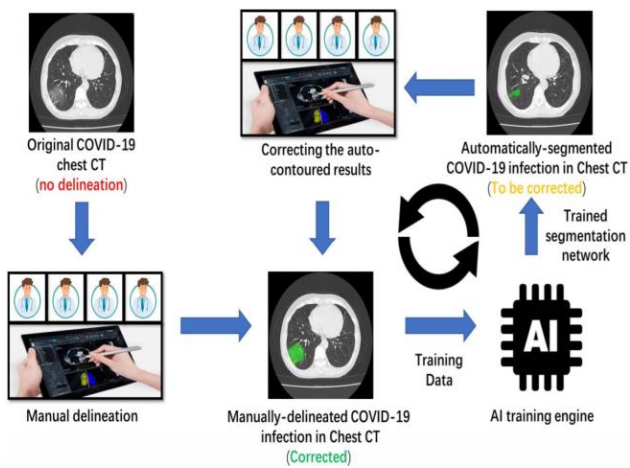

Fig. 2. Human-in-the-loop workflow in [10].

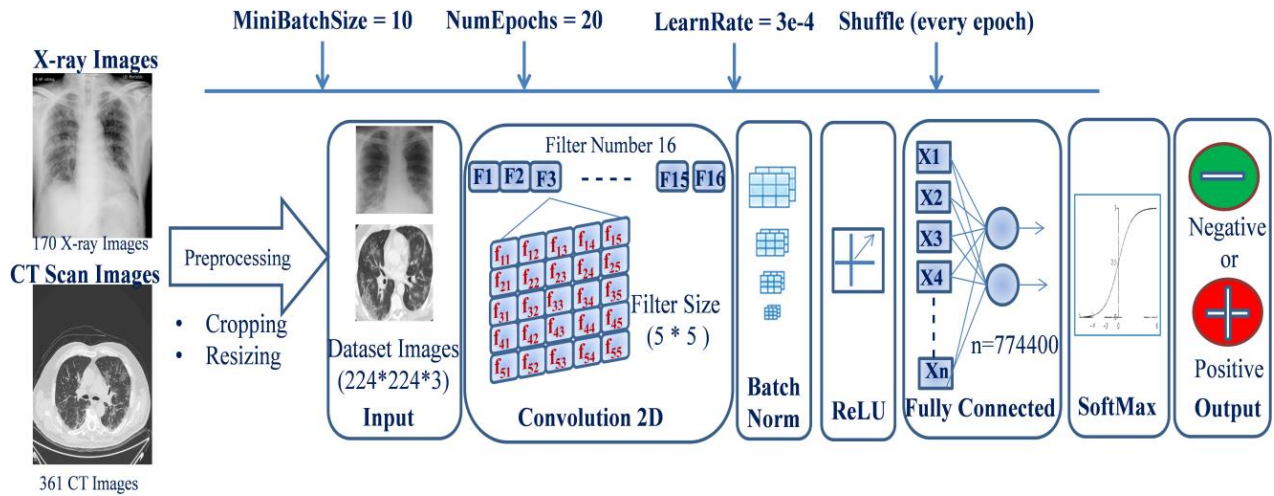

Fig. 3. The modified CNN architecture to diagnose disease COVID-19.

**Input layer:** This layer is responsible to read a pre-processed image dataset. This means, a pre-processing step is applied on the X-ray and CT scan images, separately. The pre-processing step includes cropping and resizing the images. This is because, 1) when images are taken from medical devices, several letters, art craft and medical symbols are made on them, and 2) since the images come from different sources, their sizes will be varied. Therefore, in this study, the size of the input image is changed to 224-by-224-by-3 which is width-by-height-by-channel number. We cropped the lung and chest area such that it does not contain any writing, as much as possible, a sample of the images can be seen in fig 3.

**Convolution layer:** This layer is the essential layer of our proposed CNN model in which most of the computations will be performed. The main function of this layer is to retrieve features from the image dataset and to maintain the spatial relationship between image pixels. This is obtained via learning the retrieved features using a set of filters. In this work, a two dimension convolution relied on 16 filters where each filter is constructed based on 5 * 5 filter size. Further, the filters are moving along input images and calculating the dot product function which is known as convolved features. Practically, the CNN learns these convolved features during the training process and the convolved images are of the same size as the input image.

**Batch Normalization Layer:** This layer is a very deep neural network training technique that normalizes the convolved feature values. The main reason of using this layer is to reduce the number of training epochs which are required for training deep network and stabilizing the learning process.

**ReUL layer:** The purpose of using such layer is replacing negative pixel values by zero in the convolved features. This is to produce the none-linearity map of the features in the CNN network.

**Fully Connected Layer:** The neurons at this layer are connected to all the activation functions from the previous layer. In this study, the main responsibility of this layer is to classify the retrieved convolved features from the image datasets into the defined classes.

**Softmax Layer:** This layer is simply used to interpret the probabilities values of the activation function results from the previous layer. In the diagnose disease cases; the values could be interpreted into two classes which are '0' and '1'.

**Output Layer:** This is the final layer of the CNN model in which the result values of the previous layer could be labeled. For example, the value of '1' is labeled to COVID-19 (i.e. positive case) and the value '0' is labeled to noneCOVID-19 (i.e. normal chest X-ray or CT).

**Modified AlexNet Network:** the pre-trained AlexNet network is referred to the CNN family which has been trained over a few million images on ImageNet in the range of one thousand different classes (or objects) [1]. The main purpose of using and modifying the pre-trained AlexNet is to transfer the learned weights, bias and features to the proposed approach (i.e. diagnosing disease COVID-19). This is followed by applying these parameters with training our new input dataset, i.e. CT scan images and X-ray image dataset, separately. In addition, using such transfer learning algorithms provide insight into how one can think of design new architectures for the detection of COVID-19 cases. This is because; training a CNN network based on initializing random weights value from the scratch will take long time than tuning a pre-trained network and does not require a hug computational power especially if the dataset of interest does not contain a large number of images. However, to adopt the pre-trained network for our task, the model is modified based on replacing the last layers by intended layers. Further, our collected datasets have been used for further training network, as it can be seen in Fig 4.

The input images of the dataset are cropped and the size of the images are unified according to AlexNet model, when the width is equal to 227, height is equal to 227, and the channel color number is equal to 3. Further, a set of parameters of the modified network are configured including minimum batch size is tuned to 10, number of epochs is equal to 20, learning rate is initialized to 3e-4, shuffling is set at every epoch, and finally the validation frequency has been adjusted to 3. The input images of the dataset are cropped and the size of the images are unified according to AlexNet model, when the width is equal to 227, height is equal to 227, and the channel color number is equal to 3. Further, a set of parameters of the modified network are configured including minimum batch size is tuned to 10, number of epochs is equal to 20, learning rate is initialized to 3e-4, shuffling is set at every epoch, and finally the validation frequency has been adjusted to 3.

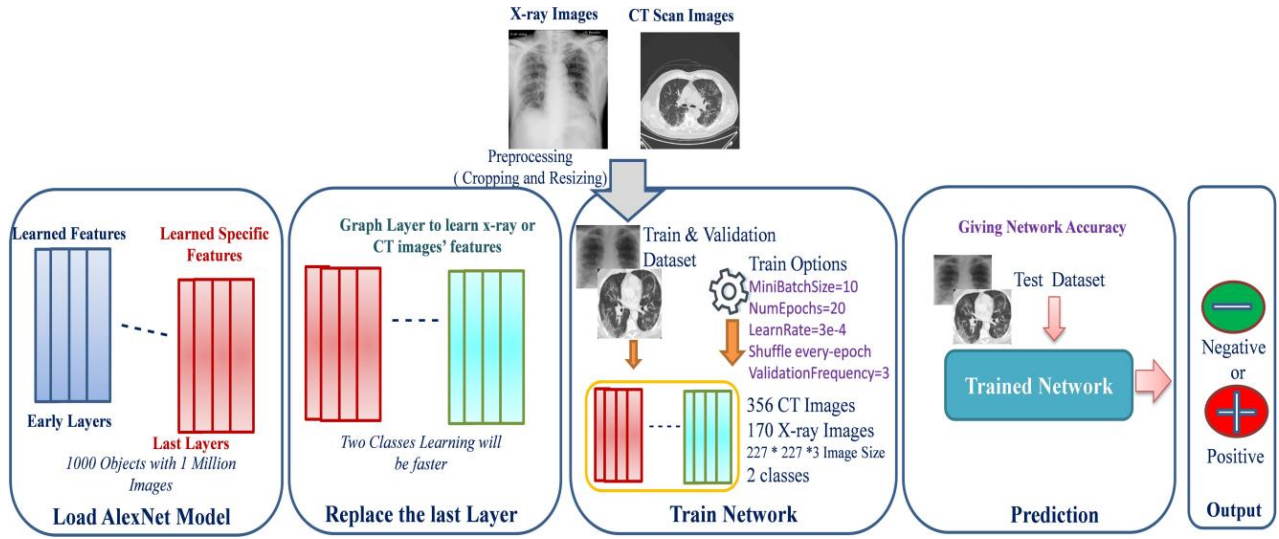

Fig. 4. The modified AlexNet network for diagnosing disease COVID-19.

## IV. MATERIALS AND METHODS

### A. Datasets

To test the proposed approach, we collected images from 5 different sources to form a dataset of 170 X-ray images and 361 CT images of COVID-19 disease. The distribution of images according to their collected sources are depicted through table 1 and table 2. There are two reasons behind using the images from these sources. First, it is diverse whereby the images collected from different sources and different countries which is important to design a sophisticated tool to help radiologists to diagnose COVID-19 around the world. Second, the images from these sources are openly available to research community and to the general public. Furthermore, the images used in this will collectively be available in a GitHub repository [20]. It is clear from table 1 that we currently do not have plenty of COVID-19 images publicly available to the research community to conduct intense investigation and there is an immediate need to collect more radiology images which can be accessible by the research community.

The X-ray images of COVID-19 are of 45 patients whereas the CT images are collected from 6 patients form British

TABLE I
DISTRIBUTION OF COVID-19 IMAGES (X-RAY AND CT) WITH RESPECT TO THEIR COLLECTED SOURCES.

| COVID-19 | GitHub | BSTI | Total |
|---|---|---|---|
| X-ray | 70 | 15 | 85 |
| CT | 16 | 187 | 203 |

TABLE II
DISTRIBUTION OF NORMAL (NO INFECTION) CHEST IMAGES (X-RAY AND CT) WITH RESPECT TO THEIR COLLECTED SOURCES.

| Normal | Kaggle | Radiopedia | Total |
|---|---|---|---|
| X-ray | 70 | 15 | 85 |
| CT | - | 153 | 153 |

Society of Thoracic Imaging (BSTI) [21] dataset and 16 patients from GitHub, in which part of the images are collected from Societa Italiana di Radiologia Medica e Interventistica (SIRM) which stands for Italian Society of Medical Radiology and Interventional [22]. Fig. 5 shows an example of the images that are used in our experiments. Both Datasets are updated regularly whereby hospitals around the world are uploading images of the disease when possible. The images are collected from both of datasets until 18th March 2020. All COVID-19 images are confirmed cases, the patient ages are varying from 35 to 84 years old and of mixed gender. At this stage there is no information about further lung infection of the patients in either of the datasets used in this study. The proposed CNN model is trained using 120 X-ray images (60 COVID19 from github, 60 normal from Kaggle [23]) and 339 CT images (COVID19: 187 BSTI and 5 from github, normal: 147 from Kaggle). More precisely, in the training phase the datasets are divided to two categories: 50 % is used to train CNN whereas

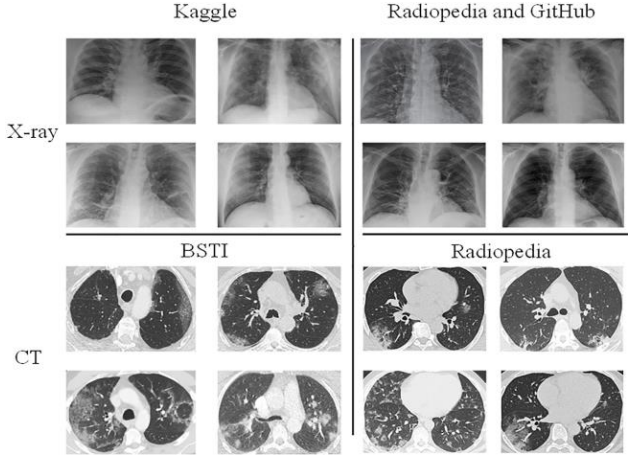

Fig. 5. CT scan and X-ray images from different sources.

50 % is used to validate the model three times in each epoch (Fig. 5).

To give credibility to the results, testing phase is adopted in the experiments. To test the proposed CNN models, we used 50 X-ray images (25 COVID-19 X-ray images from BSTI and 25 Normal X-ray images where we selected 10 images from Kaggle and 15 images from Radiopedia). For CT images, on the other hand, 17 images used in the testing phase whereby 11 of them is COVID-19 from GitHub and 6 normal CT images from radiopedia. Furthermore, it is worth mentioning that the performance measures of the experiments exhibit the average of 10 simulation runs (each run is 20 epochs with mini-batch of size 10). Note that the majority of images used in testing phase are from a different source which comes from a different device setting and a different country. The performance measures used in this study are accuracy (Acc), sensitivity and specificity which will be defined next.

$$Acc = \frac{True\ positive\ (TP) + True\ negative(TN)}{Total\ number\ of\ tested\ images} \quad (1)$$

where true positive is the number of truly identified COVID-19, true negative is the number of truly identified non-COVID-19 (normal) cases.

$$sensitivity = \frac{TP}{TP + False\ negaitve\ (FN)} \quad (2)$$

where FN is the number of COVID-19 images (X-ray or CT) incorrectly classified as non-COVID-19 and FP is the number of normal (X-ray or CT) images misclassified as COVID-19.

$$specificity = \frac{TN}{TN + False\ positive\ (FP)} \quad (3)$$

### B. Experiments & Results

This section contains the results obtained following the setup discussed in the previous section. Our proposed CNN architecture performance summarized in table 1. It is clear from the table that our proposed, rather simple, CNN method performs well on both types of images used in our inves-tigation which are X-ray and CT. Interestingly, we achieve 100 % sensitivity when X-ray is used to test the suspected patients for COVID-19 and 90 % sensitivity achieved when CT images used. Pretrained AlexNet, on the other hand, is not performing bad on X-ray images whereby it correctly identified all COVID-19 images and correctly identified 96 % of the normal X-ray images. But this is not the case when pretrained AlexNet used to differentiate COVID-19 CT images from their normal counterparts such that it only classified 72 % of the COVID-19 CT images correctly.

TABLE III
PERFORMANCE OF PROPOSED CNN AND ALEXNET FOR COVID-19 DETECTION.

| Method | Image type | Sensitivity | Specificity | Acc |
|---|---|---|---|---|
| Our CNN | X-ray | 100 | 88 | 94 |
|  | CT | 90 | 100 | 94.1 |
| AlexNet | X-ray | 100 | 96 | 98 |
|  | CT | 72 | 100 | 82 |

In comparison with a newly published CNN architecture in [1], where the same dataset from GitHub and Kaggle used, the CNN method in [1] achieved the same sensitivity as ours on COVID-19 X-ray chest images while our proposed CNN sensitivity is better than [1] for Normal X-ray images. Furthermore, there is no hint whether the CNN method proposed in [1] will work on CT images or not, whereas the CNN proposed in this study shows promising results to be used to detect COVID-19 despite the small number of images used in the current investigation. Finally, it should be noted that the CNN model proposed in this work is a very simple, yet effective, architecture which is main pillar is of only one convolutional layer which constitutes of 16 filters of size 5-by-5 trained from scratch using the weight glorot wright-initializer, see Fig. 1.

Pretrained AlexNet, on the other hand, performed well to differentiate COVID-19 X-ray images from normal X-rays but not as good as X-ray scans of chest radiographs when tested on CT images of COVID-19 patients. The performance of Pretrained AlexNet on COVID-19 and normal X-rays is illustrated in Fig. 6 and 7. As it is shown in Fig. 6, the validation accuracy is 100 % in 20 epoch and 120 iterations. This is due to the fact that the Pretrained AlexNet is modified by replacing last layers with intended layers. Further, training the Pretrained AlexNet is not adopted from scratch as conventional CNN. Pretrained AlexNet transfer learning algorithm exhibits a great performance on extracting silent features of the COVID-19 X-ray images (Fig. 8). As depicted in Fig. 8, the improved AlexNet diagnoses all COVID-19 X-ray images whereas it could correctly identified 23 normal X-ray images out of 25 images, this means the overall accuracy is around 98 %. The results show the efficiency of AlexNet in detection of lesions and opacities of infected COVID-19 patients. This will tremendously assist the radiologists by overcoming load on the medical system and hospitals.

## V. CONCLUSION

In this study, we introduced a simple yet an effective CNN model together with testing pre-trained AlexNet for the detec-tion of COVID-19 disease from chest X-ray and CT images that is available publicly. Furthermore, the images (X-ray and CT) used in our investigation are collected from multiple sources and we collectively make the radiography images used in this study to detect COVID-19 will be publicly available for the research community. Although we achieved rather high COVID-19 detection accuracy, sensitivity and specificity but this does not mean a production ready solution especially with the limited number of images currently available about COVID-19 cases. The purpose of this study is to provide radiologists, data scientists and research community with a simple CNN model which can be adopted for early diagnosis of COVID-19 and hopefully this will be built upon to accelerate the research in this direction. Also, further efforts are required to develop a system to deploy and run the proposed models on current powerful technologies, such as smartphones and Tablets. This is to help the radiologist to use the system which can be deployed on cell phones especially nowadays majority of people have smartphones with good computational power. Hence, if one can build a simple CNN model and its effective then it is easy, and does not require a lot of computational power, for testing prospective radiography images and can also be adapted easily on smartphones too. In future, we plan to collect more X-ray and CT images to expand the investigation presented in this study and if needed design a deeper CNN accordingly. We hope that the results presented in this study serves as a small step towards building a sophisticated COVID-19 disease detection from X-ray or Ct images sooner than later to save as many live as possible. Although the accuracy of the proposed models is not enough, but the result of the detection could be compensated by using other symptoms of the disease.

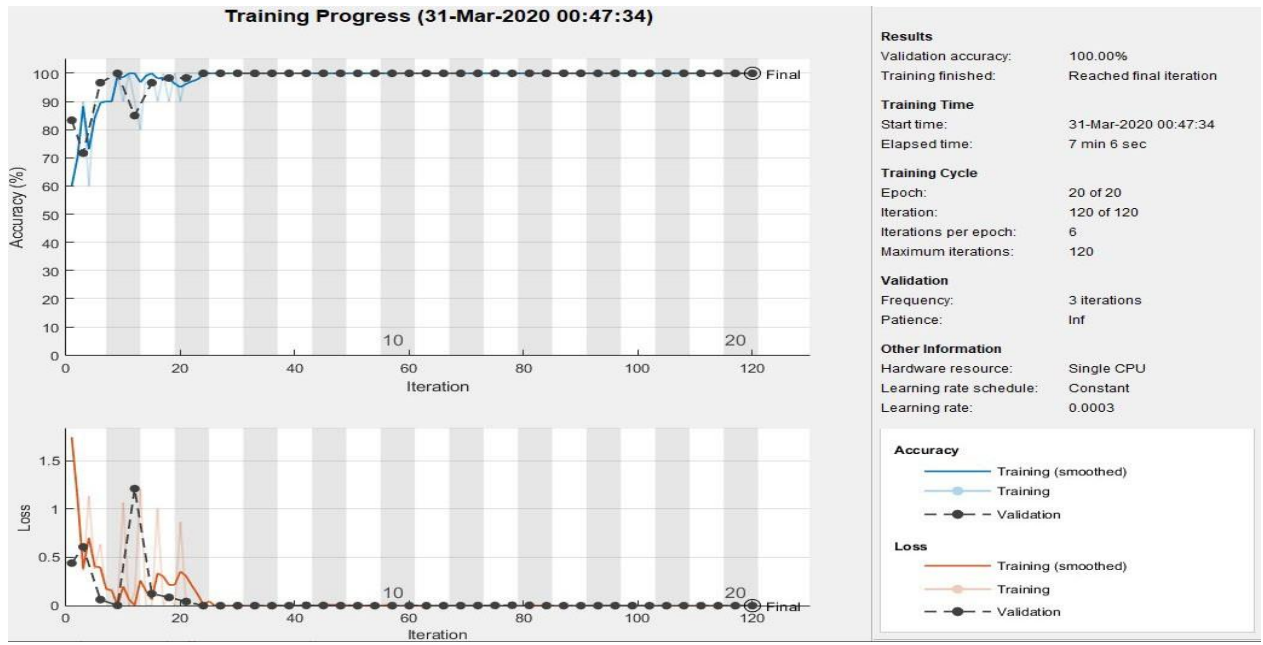

Fig. 6. Validation accuracy with respect to the 120 iterations.

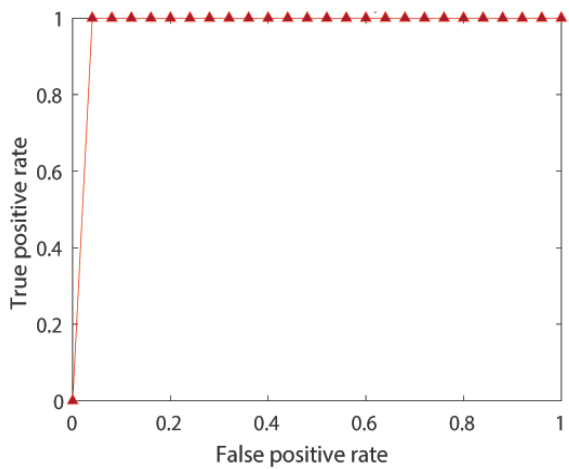

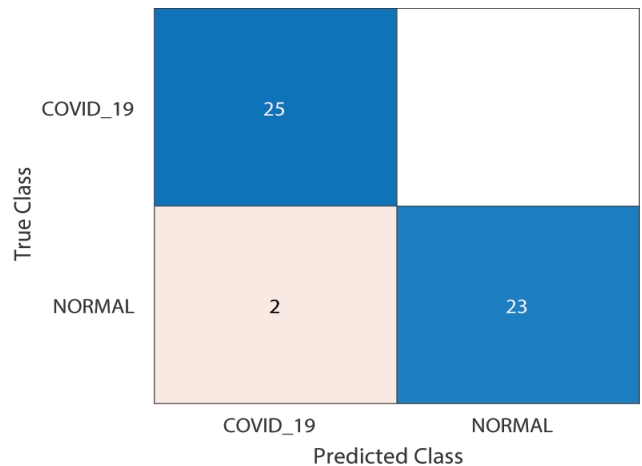

Fig. 7. True positive rate with respect to the false positive rate.

Fig. 8. Identified all COVID-19 X-ray images correctly, while 23 normal X-rays images out of 25 images.

The symptoms' level could be clearly identified via reading sensing data of onboard smartphones sensors including fever symptom level via temperature sensor, or fatigue symptom via inertial sensor, or cough symptom via microphone sensor [24].